# Self-sovereign identity as a tool for digital democracy

Roberta Centonze
*agrathaer GmbH*
Muencheberg, Germany
roberta.centonze@agrathaer.de

Roberto Reale
*Eutopian*
Rome, Italy
ORCID 0000-0001-7235-8843

*Abstract*—The importance of digital identity as a foundation for digital public services is considered. As the classical, centralised model digital identity has proven to be subject to several limitations, self-sovereign identities are proposed as replacement, especially in the context of e-government platforms and direct participation to policymaking (e.g. through e-voting tools).

## I. From centralised digital identity to SSI

Defining, representing, and managing digital identity (DI) is the unavoidable foundation stone for every conceivable idea of digital society. Each individual is *identified* through a (unique) set of *identifying* information. Such information is usually managed, following a long-established tradition, by some government or administrative authority.

However, the current model of digital identity, which from a normative standpoint is defined by the eIDAS Regulation in the EU space, suffers from a number of limitations:

- although it recognises and even encourages a federated approach, it is based on the concept of qualifying authorities, namely, of (usually governmental) agencies by which identities are declared as such
- both identities and services based thereupon are a preserve of public entities, whilst it would be desirable to foster a larger adoption of public digital identities as a foundation for commercial services as well
- it is hard to generalise digital identities to user- or community-defined classes of (non-human) agencies, such as algorithms, processes, machines, businesses, whole ecosystems
- end-users have little or no control on the amount of "identity" they disclose to service providers; most of the time, the choice is merely between either disclosing the whole of the data the service provider requires or not accessing the service at all.

For a number of reasons, amongst which the fact that the sheer figures of digital identities is factored in by DESI calculations plays no small role, EU member states are doing their best to boost them, e.g. by nudging people into applying for one in order to access a larger and larger slew of public services, although this does not necessarily implies a greater awareness by citizens and organisations or, for that matter, a more efficient access to the services themselves.

As a further shortcoming, the current model does not allow end users to *sell* access to their personal data. Although digital identity and personal data are not the same thing, they overlap inasmuch as personal data are the very fabric of digital identity (for natural persons); the latter can indeed be defined as a way of representing, organising, storing, retrieving and distributing personal data related to any given subject. There is a growing body of literature addressing the difficult problem of opening up the market of personal data to the data subjects themselves.

In order to address all such, self-sovereign identity (SSI) has been developed. SSI is the principle by which an individual should own and control their identity in all circumstances, as opposed to the State managing it on their behalf, thereby allowing everyone to get back the freedom to decide *exactly* which amount of information to disclose in any transaction with a third party, including authorities themselves. GDPR-compliant, censorship-resistant SSI is today within reach not only from a technological standpoint (by leveraging distributed ledger technology and zero-knowledge proofs) but also legally feasible.

## II. SSI as a tool for digital democracy

As innovation in democracy is more and more urgent, digital identity and even more so SSI are becoming the basic tools by which the relationship between individuals and public agencies can be reshaped. Digital platforms may be used to foster citizen involvement in public policy implementation. Institutional economy models are still underestimated in practical application to public goods governance, where stakeholders as DIs may play roles, without necessarily disclosing privacy sensitive information.

Applications of the SSI approach range from general-purpose public digital services to direct democracy platforms. With reference to the latter, without going to the extreme length of advocating some form of e-voting in general elections (which has, as far as currently-available technology is concerned, some ineliminable security flaws), it is nevertheless feasible to involve citizens in the whole decision-making chain, from proposal to impact evaluation. Governance is also a challenge in itself, even more so at the urban scale, as it requires transitioning to a distributed model in which data-driven-ness and involvement of all social parties (through citizenship platform, transparency movements, participatory budgeting, nudging, and so on) become pivotal.

In the process, it is of the utmost importance to avoid what, following George Ritzer, we could nickname the

McDonaldisation of public services: namely, prioritising efficiency, cost-effectiveness, calculability, predictability and standardisation of the public service over the capacity to coping with *individual needs on an individual basis*.

Notwithstanding the SSI being a technology-agnostic concept, a specific implementation will also be described, leveraging on work already being conducted e.g. by Sovrin.

In such implementation, digital identity is represented as a set of claims. All claims exist in the digital wallet of the data subject, each signed by the emitting party or by the subject themself, who may decide on a case-by-case basis the amount to be disclosed. A distributed ledger architecture is used to store only the public cryptographic keys (keyring), so as to ensure GDPR-compliance.

Zero-knowledge proof methods are used to ensure verifiability of claims by third parties without access to the actual claim content.

E.g., a vaccination or immunity certificate may be stored in the user wallet and only the mere existence of it (without actually disclosing detailed sanitary data) could be produced when required by an employer.

Digital identity has, however, a natural application in e-government platforms and civic engagement initiatives as a means to ensure that users are adequately represented and to protect against frauds.

When such platforms involve voting, it is necessary to ensure that a given set of requirements be adopted. Guaranteeing at the same time certainty (and legal admissibility) of the voter's identity, public verifiability (through transparency and/or audit trails), security, and secrecy is the holy grail of electronic voting. To address this point, a number of cryptographic techniques have been proposed.

Usability and good user experience are also pivotal, inasmuch as the adoption of e-government platforms, unless enforced by law (which we strongly advice against), is proven to be strongly correlated to their ability to adapt to the users usage patterns and digital skill level.

REFERENCES

[1] Regulation (EU) No 910/2014 of the European Parliament and of the Council of 23 July 2014 on electronic identification and trust services for electronic transactions in the internal market and repealing Directive 1999/93/EC

[2] Acquisti, Alessandro and Taylor, Curtis R. and Wagman, Liad, The Economics of Privacy (March 8, 2016). Journal of Economic Literature, Vol. 52, No. 2, 2016, Sloan Foundation Economics Research Paper No. 2580411

[3] The European Union Blockchain Observatory & Forum, Thematic report on blockchain and digital identity